%% file: aipsamp.tex
\documentclass{article}
\usepackage{arxiv}
\usepackage{amsmath,amssymb}
\usepackage{graphicx}
\usepackage{dcolumn}
\usepackage{bm}

\usepackage[utf8]{inputenc}
\usepackage[T1]{fontenc}
\usepackage{mathptmx}
\usepackage{color}
\newcommand{\blue}{\textcolor{blue}}

\usepackage{url}
\usepackage[hidelinks,colorlinks=true, citecolor = blue]{hyperref}
\usepackage{natbib}
\newcommand{\ssr}{    {Space Sci. Rev. }}

\newcommand{\jgr}{    {J. Geophys. Res.}}

\newcommand{\grl}{    {Geophys. Res. Lett.}}
\newcommand{\physscr}{    {Phys. Scr.}}

\newcommand{\apjss}{    {Astrophys. J .}}
\newcommand{\prlss}{    {Phys. Rev. Lett.}}
\newcommand{\apjl}{    {Astrophys. J. Lett.}}
\renewcommand{\blue}{\textcolor{black}}

\def\XXint#1#2#3{{\setbox0=\hbox{$#1{#2#3}{\int}$}
		\vcenter{\hbox{$#2#3$}}\kern-.5\wd0}}

\title{The dynamics of electron holes in current sheets\thanks{The following article has been submitted to Physics of Plasmas. After it is published, it will be found at https://publishing.aip.org/resources/librarians/products/journals/}}

\author{
Pavel Shustov\\
p.shustov@iki.rssi.ru\\
Space Research Institute of Russian Academy of Sciences, 117997, Moscow, Russia\\
Faculty of Physics, National Research University Higher School of Economics, \\
21/4 Staraya Basmannaya Ulitsa, Moscow, 105066, Russia
\And
Ilya V. Kuzichev\\
kuzicheviv@gmail.com\\
New Jersey Institute of Technology, Newark, \blue{New Jersey 07102,} USA\\
Space Research Institute of Russian Academy of Sciences, 117997, Moscow, Russia
\And
Ivan Y. Vasko\\
ivan.vasko@ssl.berkeley.edu\\
Space Sciences Laboratory, University of California at Berkeley, \blue{CA 94720,} USA\\
Space Research Institute of Russian Academy of Sciences, 117997, Moscow, Russia
\And
Anton V. Artemyev\\
aartemyev@igpp.ucla.edu \\
Institute of Geophysics and Planetary Physics, University of California, 90095, Los Angeles, USA\\
Space Research Institute of Russian Academy of Sciences, 117997, Moscow, Russia
\And
Andrew J. Gerrard\\
gerrard@njit.edu\\
New Jersey Institute of Technology, Newark, \blue{New Jersey 07102,} USA}

\begin{document}
\maketitle

\begin{abstract}
We present 1.5D Vlasov code simulations of the dynamics of electron holes in non-uniform magnetic and electric fields typical of current sheets and, particularly, of the Earth's magnetotail current sheet. The simulations show that spatial width and amplitude of electron holes do not substantially vary in the course of propagation, but there arises a double layer localized around the electron hole and manifested as a drop of the electrostatic potential along the electron hole. We demonstrate that electron holes produced around the neutral \blue{plane} of a current sheet slow down in the course of propagation toward the current sheet boundaries. The leading contribution to electron hole braking is provided by the non-uniform magnetic field, though electrostatic fields typical of the current sheets do provide a noticeable contribution. The simulations also show that electron holes with larger amplitudes are slowed faster. The simulation results \blue{suggest that some of} slow electron holes recently reported in the Earth's plasma sheet boundary layer may appear due to braking of initially fast electron holes in the course of propagation in the current sheet.
\end{abstract}

\maketitle

\section{Introduction \label{sec:intro}}


Electron phase space holes are electrostatic solitary waves produced in a nonlinear stage of various electron streaming instabilities, including electron bump-on-tail\cite{Omura96,Sigov96,Pommois17}, electron two-stream \cite{Morse69,Omura96,Goldman99}, and Buneman\cite{Drake03,Buchner06,Che10,Jara-Almonte14} instabilities. These solitary waves with a bipolar parallel electric field are purely kinetic plasma modes existing due to a dearth of the phase space density of electrons trapped by the bipolar parallel electric field \cite{Gurevich68,Turikov84,Schamel00,Hutchinson17}. Electrostatic solitary waves interpreted in terms of electron holes were measured in laboratory experiments \cite{Saeki79,Kovalenko83,Lefebvre10,Fox12} and aboard numerous spacecraft in various regions/structures of the near-Earth space environment, including reconnecting current sheets \cite{Cattell05,Graham15,Graham16}, plasma sheet boundary layer \cite{Matsumoto94,Norgren15,Lotekar20}, auroral region \cite{Ergun98,Bounds99}, inner magnetosphere \cite{Vasko15:grl,Vasko17:EHprop,Mozer15,Malaspina18}, solar wind \cite{Mangeney99,Malaspina13}, collisionless shocks \cite{Vasko18:grl,Vasko20:front,Wang20:apjl} and Lunar environment\cite{Malaspina&Hutchinson19,Chu&Halekas20}. The high-resolution plasma measurements aboard the Magnetospheric Multiscale spacecraft (MMS) have recently allowed probing the phase space density of electrons trapped in electron holes \cite{Mozer18:prl}, while multi-satellite MMS measurements allowed resolving the three-dimensional electron hole structure \cite{Tong18,Holmes18,Steinvall19}. The studies of electron holes are stimulated by a potentially substantial contribution of these solitary waves into electron heating in reconnecting current sheets that has been revealed in numerical simulations\cite{Drake03,Che10,Hesse18:separ} and spacecraft measurements\cite{Mozer16:prl,Norgren20,Khotyaintsev20:prl}. In addition, electron holes can efficiently pitch-angle scatter electrons in the inner magnetosphere resulting in electron losses to the ionosphere \cite{Vasko17,Vasko18,Shen20} and may contribute to large-scale parallel potential drops in the auroral region \cite{Mozer97,Ergun98:grl,Kuzichev17:grl}, anomalous resistivity in reconnecting current sheets\cite{Drake03,Buchner06,Che17}, and electron heating in collisionless shocks \cite{Hoshino02,Shimada&Hoshino04}. 

In terms of valuable applications, electron holes can be potentially exploited as tracers of magnetic reconnection\cite{Goldman08,Lapenta11:grl} and plasma instabilities operating on time scales not resolved by spacecraft plasma measurements \cite{Khotyaintsev10,Lotekar20}. In particular, spacecraft measurements around reconnecting current sheets in the Earth's magnetotail\cite{Cattell05,Norgren15,Lotekar20} and at the magnetopause\cite{Cattell02,Graham15,Graham16} show the presence of electron holes with distinctly different velocities. Fast electron holes propagating with velocities of a few tenths of a local electron thermal velocity are considered to be the evidence of electron bump-on-tail instabilities\cite{Graham15,Lotekar20}, while slow electron holes propagating with velocities of the order of a local ion thermal velocity are considered to be the evidence of electron two-stream and Buneman instabilities\cite{Khotyaintsev10,Norgren15}. There is a caveat though in using electron hole parameters to infer the nature of instabilities resulting in formation of the electron holes. In the case of a sufficiently long lifetime, electron holes observed aboard a spacecraft might be generated not locally, but propagated to the spacecraft from a distant generation region. This scenario might be realistic because there are reports of electron holes measured sequentially aboard several spacecraft separated in space by a few tens of kilometers \cite{Pickett04:npg,Norgren15,Lotekar20}. Since electron holes typically propagate in a plasma with parallel electric field and non-uniform magnetic field distributions, electron hole parameters can in principle substantially evolve in the course of propagation. Therefore, the analysis of the electron hole dynamics in a non-uniform plasma is of practical importance in space plasma physics.

The numerical analysis of the electron hole dynamics in a non-uniform plasma is currently restricted to a single Particle-In-Cell simulation\cite{Mandrake00} and several Vlasov simulations \cite{Briand08,Kuzichev17:grl,Vasko17:pop}. These simulations demonstrated that electron hole parameters may indeed substantially evolve in the course of propagation in a non-uniform plasma. The previous studies were restricted to simulations of the electron hole dynamics in a plasma with either non-uniform magnetic field \cite{Kuzichev17:grl} or plasma density\cite{Mandrake00,Briand08,Vasko17:pop} distributions, where in the latter case the non-uniform plasma density distribution was supported by an external potential field. The recent spacecraft measurements of electron holes with distinctly different velocities in the Earth's magnetotail\cite{Norgren15,Lotekar20} and at the magnetopause\cite{Graham15,Graham16} stimulate analysis of the electron hole dynamics in current sheets, which typical structure includes both parallel electric field and non-uniform magnetic field  distributions\cite{Zelenyi11,Artemyev11:jgr_Efield,Artemyev16:pop_anis}.

In this paper we address the dynamics of electron holes in a non-uniform plasma typical of the Earth's magnetotail current sheet using 1.5D Vlasov code originally developed and used for analysis of the electron hole dynamics in non-uniform magnetic fields\cite{Kuzichev17:grl}. We consider the evolution of electron hole parameters and discuss the results in the light of recent spacecraft measurements of slow electron holes in current sheets in the Earth's magnetosphere. We formulate the problem in Section \ref{sec1}, describe the numerical scheme of the Vlasov code in Section \ref{sec2}, present simulation results in Section \ref{sec3}, and discuss and summarize the results in Section \ref{sec4}.

\section{Problem formulation and simulation setup\label{sec1}}

\subsection{Background magnetic field and plasma distributions}

\begin{figure*}[ht!]
    \centering
    \includegraphics[width=0.8\textwidth]{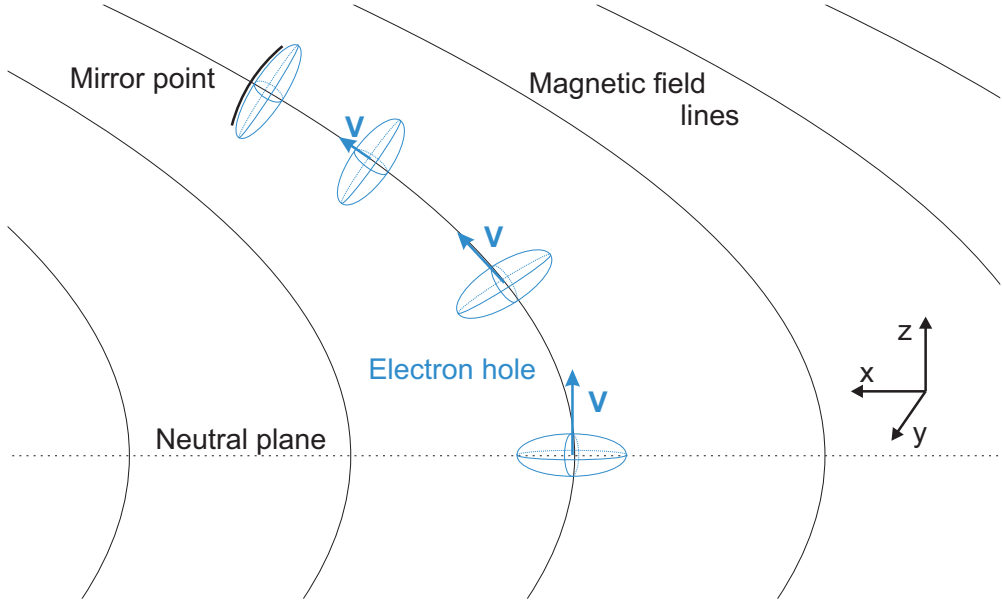}
    \caption{\blue{Schematics of an electron hole propagation in the Earth's magnetotail current sheet ($xyz$ is the Geocentric Solar Ecliptic coordinate system with the $x$-axis pointing toward the Sun, the $z$-axis perpendicular to the ecliptic plane and the $y$-axis completing the right-hand coordinate system). The magnetic field lines have typical slingshot configuration in the $xz$-plane. The electron hole is produced by some instability around the neutral plane, where the magnetic field has minimum value, and propagates toward the current sheet boundary along a non-uniform magnetic field. Electron holes are generally three-dimensional structures\cite{Franz00,Tong18}, but in the Earth's magnetotail electron holes are approximately one-dimensional, because the perpendicular spatial scales are one-two orders of magnitude larger than parallel spatial scales \cite{Cattell05,Lotekar20}. Both parallel and perpendicular scales are much smaller than typical spatial scale of the magnetic field variation that is of a few thermal ion gyroradii. In the simulations, we consider one-dimensional electron hole propagating along a particular magnetic field line and the only spatial coordinate $s$ along the magnetic field line is measured from the neutral plane. We assume electrons to be magnetized and include effects of perpendicular electron dynamics through considering electrons with different magnetic moments (Section \ref{sec1}).}}
    \label{fig0}
\end{figure*}

\blue{Figure \ref{fig0} presents a schematics of electron hole propagation to the boundary of the current sheet from the neutral plane, where the electron holes can be produced by various instabilities (e.g., electron bump-on-tail or electron two-stream instability). The measurements in the Earth's magnetotail current sheet showed the presence of fast electron holes, which are most likely produced by electron bump-on-tail instability \cite{Holmes18,Tong18,Lotekar20}. In the Earth's magnetotail current sheet, the electron plasma frequency is typically a few orders of magnitude larger than the electron cyclotron frequency and electron holes typically have perpendicular spatial scales much larger than parallel spatial scales\cite{Cattell05,Lotekar20}, because the perpendicular scales are of a few thermal electron gyroradii, while parallel spatial scales are of a few Debye lengths \cite{Franz00}. The spatial scales of electron holes are several orders of magnitude smaller than a typical thermal ion gyroradius that is a characteristic spatial scale of the magnetic field variation in the Earth's magnetotail current sheet\cite{Petrukovich15:ssr}.} Because electron holes are approximately one-dimensional and \blue{small-scale} structures propagating \blue{(according to spacecraft measurements)} along magnetic field lines\cite{Cattell05,Norgren15,Lotekar20}, it is sufficient to consider the dynamics of electron holes propagating along a particular magnetic field line. In the Earth's magnetotail current sheet, the magnetic field along magnetic field lines is non-uniform, while the plasma density has much weaker field-aligned gradients and can be assumed more or less uniform \cite{Zelenyi11,Petrukovich15:ssr}. The electron population is magnetized and has some temperature anisotropy, which results in the appearance of electrostatic fields parallel to magnetic field lines \cite{Artemyev11:jgr_Efield,Artemyev16:pop_anis}. Thus electron holes propagate in a non-uniform magnetic field $B(s)$ and electrostatic potential $\varphi(s)$, where $s$ is the coordinate along the magnetic field line measured from the neutral \blue{plane}. The magnetic field has minimum value in the neutral \blue{plane} and becomes uniform at sufficiently large distances from the neutral \blue{plane\cite{Petrukovich15:ssr}}.

We first describe equilibrium distributions of magnetic and electrostatic fields and electron velocity distribution functions (VDF) along the magnetic field line. \blue{The electron VDF in the neutral plane is assumed to be anisotropic Maxwell}
\begin{eqnarray*}
f_{e}({\bf v},s=0)=A_0\;n_0\;\left(\frac{m_{e}}{2\pi T_{||}}\right)^{3/2}\exp\left[-\frac{m_{e}v_{||}^2}{2T_{||}}-A_0\frac{m_{e}v_{\perp}^2}{2T_{||}}\right],
\end{eqnarray*}
\blue{where $v_{||}$ and $v_{\perp}$ are electron velocities parallel and perpendicular to the magnetic field, $T_{||}$ is the parallel electron temperature, $A_0=T_{||}/T_{\perp}$ is the electron temperature anisotropy in the neutral plane, $m_{e}$ is the electron mass.} The electron VDF can be determined at every point along the magnetic field line by solving the stationary Vlasov equation \cite{book_Schindler07}. Because electrons are magnetized, magnetic moment \blue{$\mu=m_{e}v_{\perp}^2/2B(s)$ and total energy $W = m_{e}v_{||}^2/2 + \mu B(s) - e\varphi(s)$ are integrals of motion ($e$ is the elementary charge)}, one can express the electron VDF in the neutral \blue{plane} as a function of $W$ and $\mu$ 
\begin{eqnarray*}
\blue{f_{e}(W,\mu,s=0)=A_0\;n_0\;\left(\frac{m_{e}}{2\pi T_{||}}\right)^{3/2}\exp\left[-\frac{W+e\varphi_0+(A_0-1)\;\mu\; B_0}{T_{||}}\right],}
\end{eqnarray*}
\blue{where $B_0$ and $\varphi_0$ are magnetic field and electrostatic potential in the neutral plane (without loss of generality, we set $\varphi_0=0$).} We use the Liouville theorem to determine the electron VDF at every point along the magnetic field line\cite{Whipple91,Artemyev16:pop_anis}
\begin{eqnarray*}
\blue{f_{e}(v_{||},\mu,s)=A_0\;n_0\;\left(\frac{m_{e}}{2\pi T_{||}}\right)^{3/2}\exp\left[-\frac{m_{e}v_{||}^2}{2T_{||}}+\frac{e\varphi(s)}{T_{||}}-A(s)\frac{\mu B(s)}{T_{||}}\right],
\label{eq:vdf_phys}}
\end{eqnarray*}
\blue{where $A(s)$ is the position-dependent temperature anisotropy}
\begin{eqnarray*}
\blue{A(s)=1+(A_0-1)B_0/B(s)}
\end{eqnarray*}
\blue{The temperature anisotropy $A(s)$ equals $A_0$ in the neutral plane and decreases toward the current sheet boundaries, i.e. as $B(s)$ increases, $A(s)$ becomes closer to 1. We also note that the parallel temperature remains constant along the magnetic field line. The electron density is calculated as follows}
\begin{eqnarray*}
    \blue{n_e(s)= \blue{2\pi}\int f_e(v_{||},\mu,s)B(s)dv_{||}d\mu=e^{e\varphi(s)/T_{||}}A_0/A(s)
    \label{eq:ne}}
\end{eqnarray*} 
\blue{Because the plasma density along the magnetic field line is assumed to be uniform, there should be an electrostatic field which would keep ion and electron densities approximately identical. Assuming $n_{e}(s)=1$, we obtain the electrostatic potential}
\begin{eqnarray}
    \blue{e\varphi(s)=-T_{||}\ln\left(A_0/A(s)\right)}
    \label{eq:phi_phys}
\end{eqnarray}
\blue{The physical reason for appearance of this electrostatic field is rather transparent: in the case of isotropic electrons in the neutral plane ($A_0=1$), the density of electrons at any point along the magnetic field line coincides with the density in the neutral plane and no electrostatic field is necessary to equalize
ion and electron densities. In the case of a parallel anisotropy ($A_0=T_{||}/T_{\perp}>1$), the electrostatic field is directed toward the current sheet boundaries to prevent an excess of electrons at any given point along the magnetic field line; in the case of a perpendicular anisotropy ($A_0=T_{||}/T_{\perp}>1$), the electrostatic field is directed toward the neutral plane to drag a sufficient amount of electrons to any given point along the magnetic field line.}

Eq. (\ref{eq:phi_phys}) is asymptotically correct provided that typical spatial scale of the magnetic field $B(s)$ is much larger than the Debye length, which is certainly the case in the Earth's magnetotail current sheet\cite{Zelenyi11,Petrukovich15:ssr}. More precisely, the electrostatic potential $\varphi(s)$ can be determined by solving numerically the Poisson equation \blue{$\nabla^2\varphi=4\pi e (n_e(s)-n_0)$,} which can be written as follows
\begin{eqnarray*}
B(s)\frac{\partial}{\partial s}\left[\frac{1}{B(s)}\frac{\partial \varphi}{\partial s}\right]=\blue{\blue{4\pi e\left(\frac{A_0}{A(s)}e^{e\varphi(s)/T_{||}}-1\right)}}
\label{eq:poisson}
\end{eqnarray*}
with initial conditions $\varphi(0)=0$ and $\left(\partial \varphi/\partial s\right)_{s=0}=0$. We recall that $s$ is the coordinate along the magnetic field line, which results in appearance of the magnetic flux tube cross section $1/B(s)$ on the left-hand side of Eq. (\ref{eq:poisson}). Although the result of the numerical integration of the Poisson equation differs from the asymptotic expression (\ref{eq:phi_phys}) by a negligible amount (less than percent in our simulations), the use of the precise expression for $\varphi(s)$ allows excluding small-amplitude Langmuir oscillations, which are present in the simulations with $\varphi(s)$ determined by Eq. (\ref{eq:phi_phys}), because of a slight charge density imbalance at the initial moment.

In what follows, we use normalized quantities and, particularly, the magnetic field $B(s)$ normalized to its minimum value in the neutral \blue{plane} $B_0$, electron density $n_{e}$ normalized to unperturbed plasma density $n_0$, electrostatic potential $\varphi(s)$ in units of \blue{$T_{||}/e$}, spatial coordinate $s$ in units of Debye length $\lambda_{D}=\left(T_{||}/4\pi e^2n_0\right)^{1/2}$, electron velocities in units of thermal velocity $\left(T_{||}/m_e\right)^{1/2}$ and time in units of electron plasma frequency $\omega_p=(4\pi n_0e^2/m_e)^{1/2}$. \blue{In normalized units, the electron VDF at any point along the magnetic field line is written as follows}
\begin{eqnarray}
    f_e(v_{||},\mu,s)= A_0\;(2\pi)^{-3/2}\;e^{-v_{||}^2/2+\varphi(s)-\blue{A(s)}\mu B(s)}
    \label{eq:vdf}
\end{eqnarray}
where
\begin{eqnarray}
    \blue{\mu=v_{\perp}^2/2B(s),\;\;\;A(s)=1+(A_0-1)/B(s),\;\;\;
    \varphi(s)\approx -\ln(A_0/A(s))}
    \label{eq:all}
\end{eqnarray}
In the case of isotropic \blue{electrons in the neutral plane ($A_0=1$)}, the electron density is $n_{e}=e^{\varphi}$ and, hence, the electrostatic potential should be approximately zero to have $n_e(s)\approx 1$. The simulations of the electron hole dynamics for that case were performed by Kuzichev et al.\cite{Kuzichev17:grl}. The electron population in the Earth's magnetotail has a typical temperature anisotropy $1\lesssim A_0\lesssim 2$ (see Refs. \cite{Petrukovich15:ssr} and \cite{Artemyev20:jgr-anis}) and, hence, there is typically an electrostatic field present \blue{to ensure the plasma quasi-neutrality, as discussed above}. The goal of this study is to include this electrostatic field into numerical simulations of the electron hole evolution.

\begin{figure*}[ht!]
    \centering
    \includegraphics[width=1.0\textwidth]{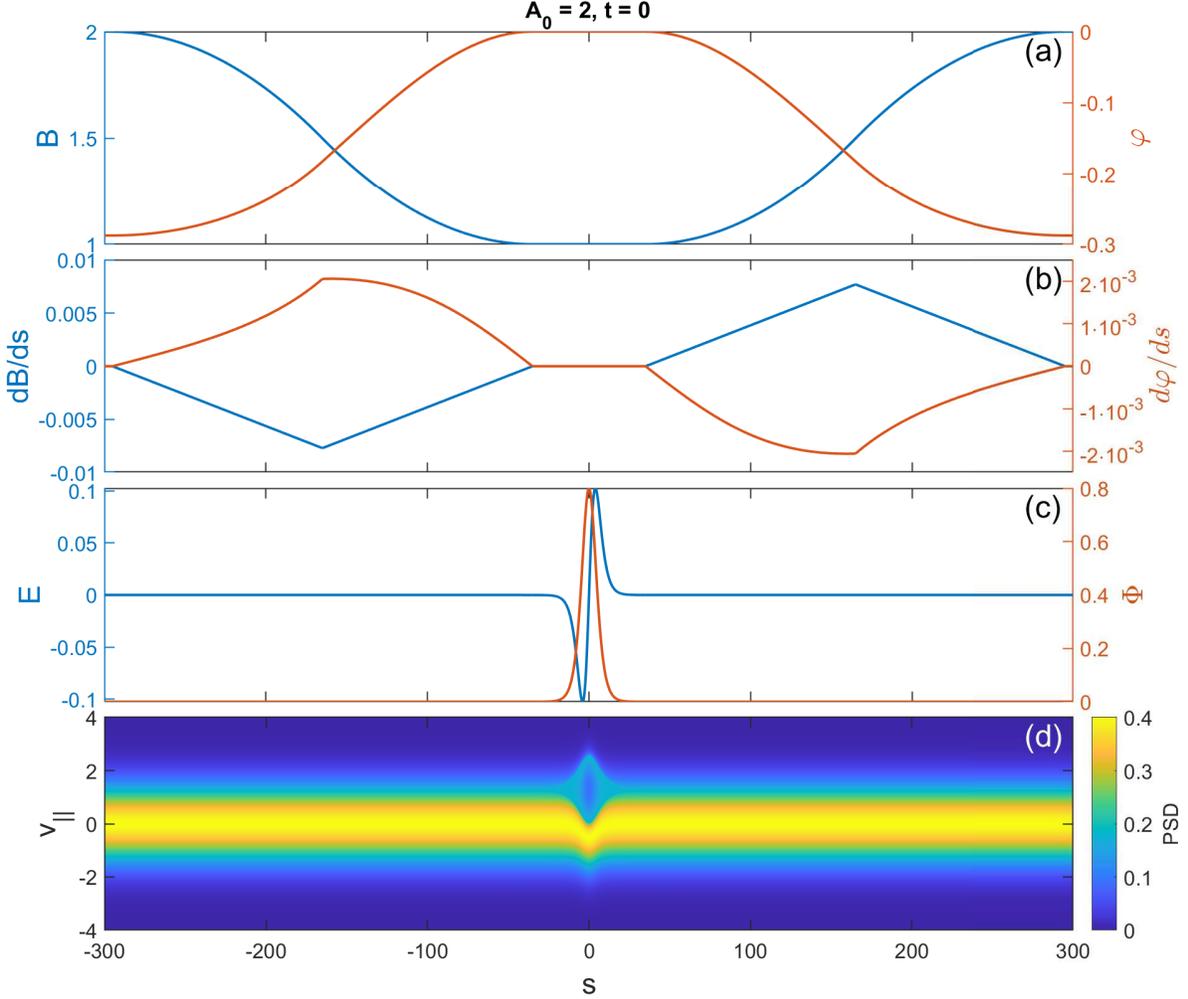}
    \caption{Setup of 1.5D Vlasov simulation of the electron hole dynamics in the current sheet: (a) magnetic field $B(s)$ and global electrostatic potential $\varphi(s)$ corresponding to electron temperature anisotropy $A_0=2$; (b) magnetic field gradient $\partial B/\partial s$ and electrostatic potential gradient $\partial \varphi/\partial s$; (c) electrostatic potential $\Phi(s)$ and parallel electric field $E_{||}=-\partial \Phi/\partial s$ of the electron hole at the initial moment; (d) reduced electron velocity distribution function at the initial moment, $F_{e}(s,v_{||})=\blue{2\pi}\int f_{e}|_{t=0}\;B(s)\;d\mu$. The electron hole is initially located in the neutral \blue{plane}, where $B(s)$ and $\varphi(s)$ are uniform, and propagates toward the right current sheet boundary.}
    \label{fig1}
\end{figure*}

Figure \ref{fig1}a demonstrates magnetic field $B(s)$ used in the simulations along with electrostatic potential $\varphi(s)$ corresponding to electron temperature anisotropy $A_0=2$. \blue{Panel (b) presents the magnetic field gradient $\partial B/\partial s$ and the electrostatic potential gradient $\partial \varphi/\partial s$.}  The simulation box is $s \in \left[-300, 300\right]$. The magnetic field is uniform around the neutral \blue{plane}, i.e. $B(s)=1$ at $|s|<s_{\rm min}=35$, and also sufficiently far from the neutral \blue{plane}, i.e. $B(s)=2$ at $|s|>s_{\rm max}=295$. In the transition region $|s|\in [s_{\rm min},s_{\rm max}]$ the magnetic field varies smoothly with continuous $\partial B/\partial s$ 
\begin{eqnarray}
    B(s) = \begin{cases}
                1+2\;\left(|s|-s_{\rm min}\right)^2/\left(s_{\rm max}-s_{\rm min}\right)^2, &s_{\rm min} < |s| \leq s_{0} \\
                2-2\;\left(|s|-s_{\rm max}\right)^2/\left(s_{\rm max}-s_{\rm min}\right)^2, &s_{0} < |s| \leq s_{\rm max}\\
           \end{cases}
           \label{eq:magnetic}
\end{eqnarray}
where $s_{0}=\left(s_{\rm max}+s_{\rm min}\right)/2=165$. In the realistic Earth's magnetotail current sheet, the magnetic field in the neutral \blue{plane} varies from 1 nT (in thin current sheets) to 10 nT (in dipolarized current sheets), the magnetic field at the current sheet boundaries is typically 10-20 nT, the current sheet thickness is of the order of 1000 km (Refs. \cite{Petrukovich15:ssr} and \cite{Vasko15:tail}), while the typical Debye length is about 1 km (Ref. \cite{Lotekar20}). In normalized units used in the simulations, $B(s)$ in the Earth's magnetotail current sheet should be within $[2,10]$ range at $|s|\gtrsim 1000$, so that $\partial B/\partial s$ is typically of about $\sim [0.01, 0.002]$. In the presented simulations the current sheet thickness is about 300, but we also assume smaller magnetic field at the current sheet boundaries, so that typical magnetic field gradient in the simulations \blue{shown in Figure \ref{fig1}b} is consistent with those in the realistic Earth's magnetotail current sheet. The same statement is valid for the electrostatic field, because Eq. (\ref{eq:all}) unequivocally relates $\varphi(s)$ to $B(s)$.

\subsection{Simulation setup of electron holes}

This study is focused on analysis of the electron hole dynamics in a non-uniform plasma and, therefore, we leave aside modelling of an electron streaming instability resulting in electron hole formation. In addition, we assume ions to be an immobile charge-neutralizing background that is justified while electron hole velocities exceed typical ion thermal velocity\cite{Schamel82,Hutchinson17}. We consider a single electron hole initially located around the neutral \blue{plane}, $|s| \leq s_{\rm min}$, where the magnetic field is uniform, $\partial B/\partial s=0$, and electrostatic field is absent, $\partial \varphi/\partial s=0$. At the initial moment, the electron hole is assumed to propagate toward the right current sheet boundary. 

Figure \ref{fig1}c presents the electric field and electrostatic potential of the electron hole. Because the electron hole is initially around the neutral \blue{plane}, where $B(s)$ and $\varphi(s)$ are uniform, the initial electron VDF consistent with the electron hole can be determined using existing approaches of calculating the Bernstein-Green-Kruskal (BGK) equilibria \cite{Bernstein57,Schamel82}. In the approach proposed by Bernstein et al.\cite{Bernstein57} and developed for electron holes by V. Turikov\cite{Turikov84}, one assumes almost arbitrary electron hole electrostatic potential and determines the VDF of electrons trapped in the electron hole. We assume the following electrostatic potential of the electron hole at the initial moment
\begin{eqnarray}
\label{eq:init_potential}
    \Phi|_{t=0} = \Phi_0\;\cosh^{-2}\left(s/D\right),
\end{eqnarray}
where $\Phi_0$ and $D$ are electron hole amplitude and spatial width respectively. 
At distances sufficiently far from the electron hole, that is at $|s|\gg D$, the electron VDF remains unperturbed
\begin{eqnarray}
f_{e}|_{t=0,|s|\gg D}=A_0(2\pi)^{\blue{-3/2}}e^{-v_{||}^2/2+\varphi(s)-\mu B(s)\blue{A(s)}}
\label{eq:bc}
\end{eqnarray}
Using this VDF as a boundary condition at $|s|\gg D$ for electrons not trapped in the electron hole, we determine the VDF of electrons not trapped in the electron hole at every point along the magnetic field line\cite{Turikov84,Kuzichev17:grl}
\begin{eqnarray}
    f_{e}|_{t=0} =A_0(2\pi)^{\blue{-3/2}}e^{-(v_H+\sigma \sqrt{w})^2/2+\varphi(s)-\mu B(s)\blue{A(s)}},
\label{eq:init_psd}
\end{eqnarray}
where $v_{H}$ is initial electron hole velocity, $\sigma = {\rm sgn}(v_{||}-v_H)$, and $w = (v_{||}-v_H)^2-2\;\Phi|_{t=0}$. Eq. (\ref{eq:init_psd}) determines only the electron VDF in the region of the phase space, where $(v_{||}-v_H)^2\geq 2\;\Phi|_{t=0}$, that is the VDF of electrons not trapped in the electron hole. Using the standard procedure described in detail by V. Turikov\cite{Turikov84}, we determine the distribution function of electrons trapped in the electron hole, that is the electron VDF in the phase space region, where $(v_{||}-v_H)^2\leq 2\;\Phi|_{t=0}$.

Fig.~\ref{fig1}d presents the reduced electron VDF, $F_{e}(s,v_{||})|_{t=0}=\blue{2\pi}\int_0^{\infty} f_{e}|_{t=0}\;B(s)d\mu$, in the entire phase space $(s,v_{||})$ at the initial moment. This VDF corresponds to the electron hole with initial amplitude $\Phi_0=0.8$, spatial width $D=6$, and velocity $v_{H}=1.3$, which electric field and electrostatic potential are shown \blue{on panel (c)}. The center of the electron hole is located at $v_{||}=v_H$ and there is a dearth of the phase space density of electrons trapped in the electron hole in accordance with the physical nature of electron holes\cite{Schamel82,Schamel00}. The reduced electron distribution function $F_{e}(s,v_{||})|_{t=0}$ is spatially uniform outside of the electron hole, because at $|s|\gg D$ we have $F_{e}(s,v_{||})|_{t=0}\approx (2\pi)^{-1/2} \exp(-v_{||}^2/2)$ and the electron density is essentially uniform, $n_{e}(s)\approx 1$.


\begin{figure*}[ht!]
    \centering
    \includegraphics[width=0.9\textwidth]{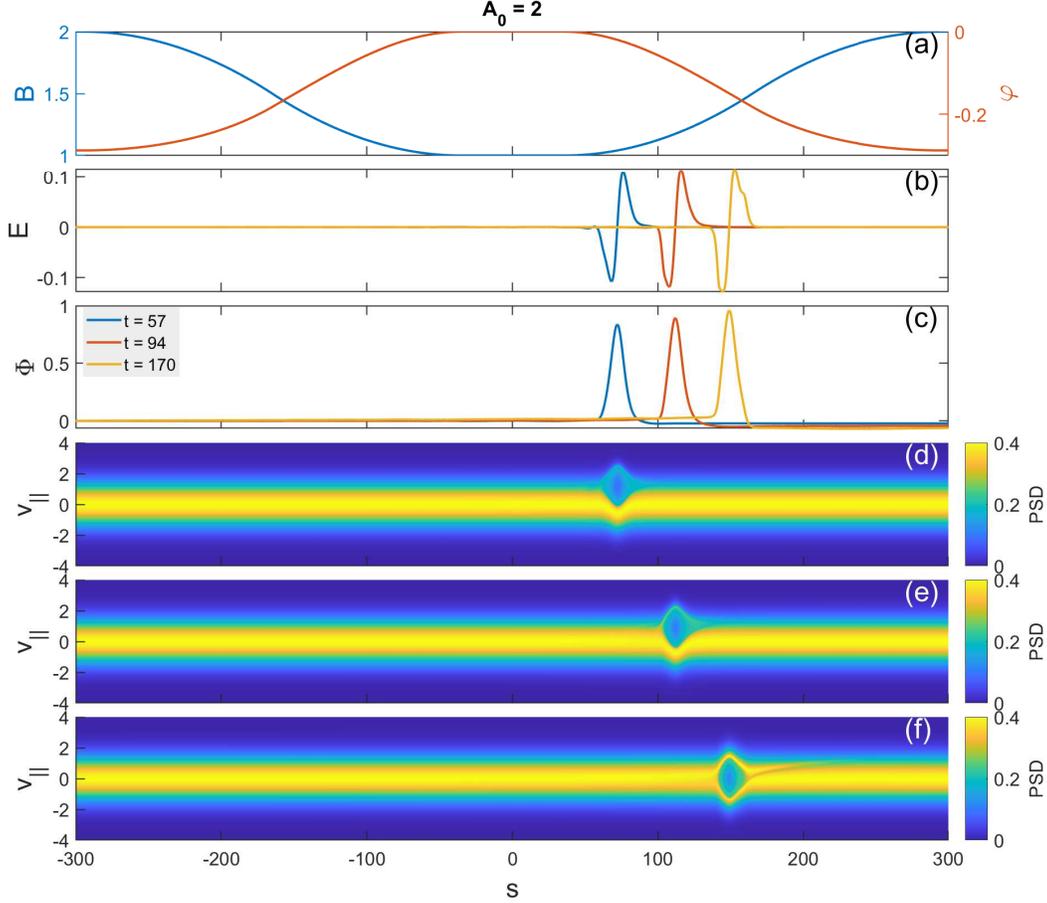}
    \caption{The results of simulation of the dynamics of an electron hole with initial amplitude $\Phi_0=0.8$, spatial width $D=6$ and velocity $v_{H}=1.3$: (a) magnetic field $B(s)$ and global electrostatic potential $\varphi(s)$ corresponding to electron temperature anisotropy $A_0=2$; (b, c) parallel electric field and electrostatic potential of the electron hole at several moments of time indicated in panel (c); (d)-(f) reduced electron distribution function $F_{e}(s,v_{||},t)=\blue{2\pi}\int f_{e}\;B(s)\;d\mu$ at the corresponding moments of time.}
    \label{fig2}
\end{figure*}

\section{1.5D Vlasov code and numerical scheme\label{sec2}}

Below we present the simulation code used to address the electron hole dynamics in the current sheet. Because electrons can be assumed magnetized, the dynamics of the electron population can be described within the gyro-kinetic approximation, so that $f_e(s, v_{||}, \mu, t)$ satisfies the following Vlasov equation\cite{Lee83}
\begin{eqnarray}
\label{eq:main}
    \frac{\partial f_e}{\partial t} + v_{||}\frac{\partial f_e}{\partial s} + \left( \frac{\partial \varphi}{\partial s} - \mu \frac{\partial B}{\partial s} - E_{||} \right) \frac{\partial f_e}{\partial v_{||}} =0
\end{eqnarray}
where $E_{||}=-\partial \Phi/\partial s$ is the parallel electric field of the electron hole. The Vlasov equation is supplemented by the Amp\'{e}re equation
\begin{eqnarray}
\label{eq:ampere}
    \frac{\partial E_{||}}{\partial t} = -j_{||} = \blue{2\pi} \int v_{||}\;f_e(s, v_{||}, \mu, t)\;B(s)dv_{||} d\mu.
\end{eqnarray}
We solve the system of Vlasov-Amp\'{e}re equations as follows. Because $\mu$ is a conserved quantity, we split the electron population into sub-populations with different values of $\mu$. The dynamics of each sub-population \blue{is} governed by the Vlasov equation (\ref{eq:main}), while the mutual interaction between the sub-populations is determined by electric field $E_{||}$ satisfying the Amp\'{e}re equation (\ref{eq:ampere}). The system of Vlasov-Amp\'{e}re equations with initial conditions (\ref{eq:init_potential}) and (\ref{eq:init_psd}) completely determines the electron hole dynamics.


The Vlasov equation for each sub-population represents a transport equation in $(s,v_{||})$ phase space, which can be solved by a splitting method (see, e.g., Ref. \cite{Valentini2005} for details). In brief, at every time step $\Delta t$, the Vlasov equation is solved by consecutive transport of the phase space density along spatial and velocity coordinates
\begin{eqnarray*}
f(t) &\longrightarrow& f^* = T_s(\Delta t/2) f(t),\nonumber\\ 
f^* &\longrightarrow& f^{**} = T_{v_{||}}(\Delta t) f^*, \\
f^{**}&\longrightarrow& f(t+\Delta t) = T_s(\Delta t/2) f^{**},\nonumber
\label{eq:SplitMeth}
\end{eqnarray*}
where $f$ stands for the electron VDF $f_{e}$ to simplify notations and $T_{s}(\tau)$ and $T_{v_{||}}(\tau)$ are operators providing transport in $s$ and $v_{||}$ directions over time interval $\tau$. The indicated order of transport operators allows obtaining the second order accuracy $\mathcal{O}(\Delta t^2)$ of the splitting method. Each transport operator is determined by the flux-balance method \cite{Fijalkov1999}, which gives the following expression, for instance, for $T_s(\Delta t/2)$
\begin{eqnarray*}
    f_{ij}^*|_{v_j \geq 0} = f_{ij} &+& \blue{\gamma}\left(f_{i-1j} - f_{ij}\right)+\nonumber\\&+&\blue{\gamma}\dfrac{\Gamma}{4}\left(f_{ij}-f_{i+1j}+f_{i-1j}-f_{i-2j} \right),\\\nonumber
    f_{ij}^*|_{v_j<0} = f_{ij} &-& \blue{\gamma}\left(f_{i+1j} - f_{ij}\right)+\nonumber\\&+&\blue{\gamma}\dfrac{\Gamma}{4} \left(f_{i+2j}-f_{ij}+f_{i-1j}-f_{i+1j}\right)\nonumber
\end{eqnarray*}
where $f_{ij} \equiv f(s_i, v_j, \mu)$ at a given moment of time, \blue{$\gamma \equiv v_j{\Delta t}/{2\Delta s}$},  $\Delta s$ is a step of the spatial grid, and $\Gamma=1-\blue{|\gamma|}$. The expression for the operator $T_{v_{||}}(\tau)$ can be written by analogy substituting spatial grid step $\Delta s$ by velocity step $\Delta v_{||}$.

In the next section, we present results of simulations for electron holes with different amplitudes propagating in current sheets with various values of electron \blue{temperature} anisotropy $A_0$. In all simulations, we use periodic boundary conditions for \blue{the} electron hole electric field and VDFs of electron sub-populations. We use integration steps \blue{$\Delta t = 0.007$, $\Delta s=0.11$, and $\Delta v_{||}=0.028$} and split the electron population into 75 sub-populations with magnetic moments $\mu$ evenly distributed in the range $[0, \mu_{\rm max}]$, where $\mu_{\rm max}\geq 9$. The specific value of $\mu_{\rm max}$ is different in simulations performed for different values of parameter $A_0$ and chosen to minimize the initial charge imbalance, which appears because of numerical integration over $\mu$ in the Amp\'{e}re equation (\ref{eq:ampere}).

\begin{figure*}[ht!]
    \centering
    \includegraphics[width=1\textwidth]{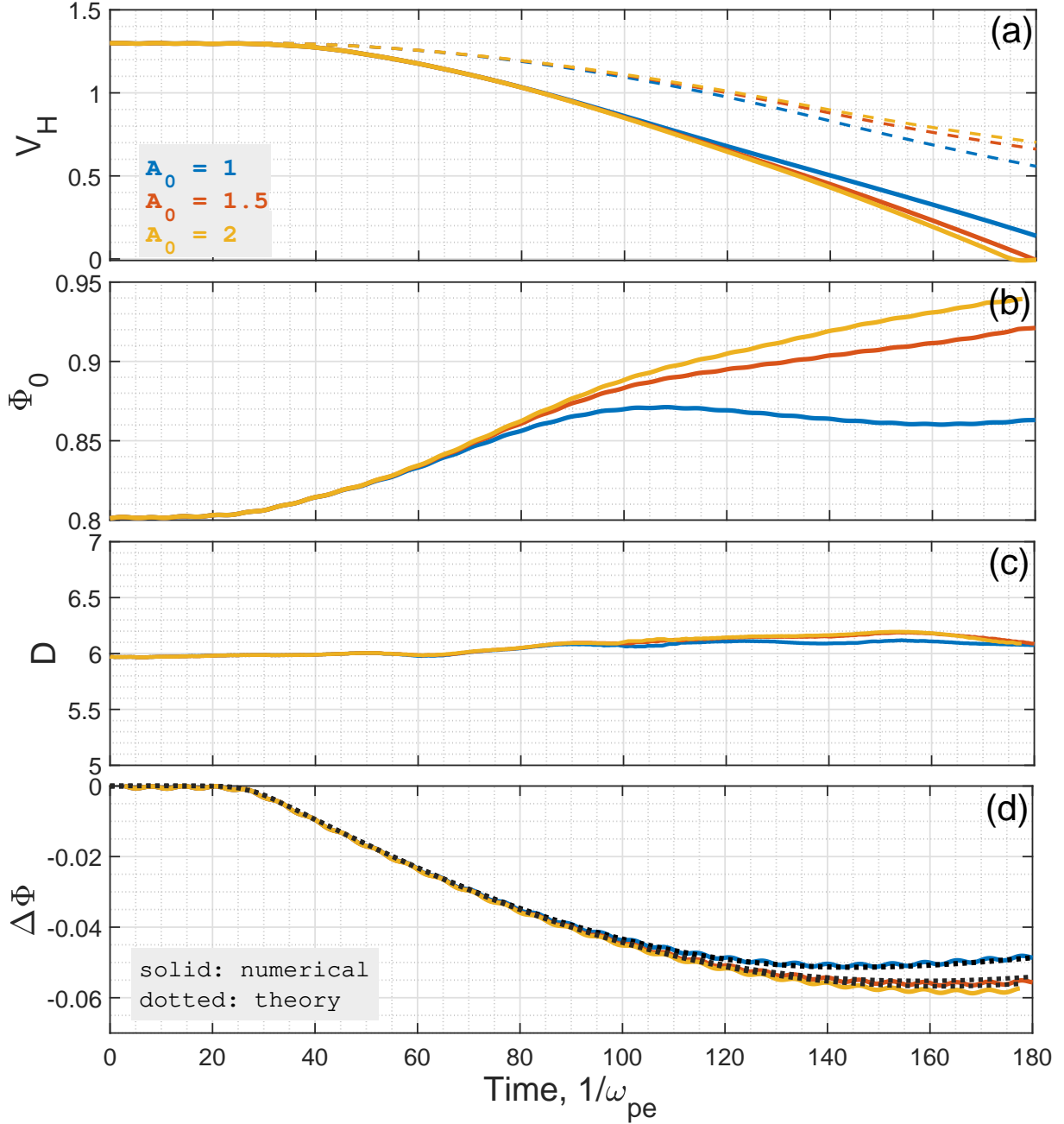}
    \caption{Summary of simulations performed for the same electron hole with initial parameters $\Phi_0=0.8$, $D=6$ and $v_{H}=1.3$, but in the current sheets with different global electrostatic fields corresponding to electron temperature anisotropies $A_0=1,$ 1.5, and 2. The panels present evolution of (a) electron hole velocities (b) amplitudes of the electrostatic potential; (c) spatial width, computed as the distance between maximum and minimum peaks of the electron hole electric field; (d) drop of \blue{the electrostatic potential} along electron holes \blue{(solid lines correspond to numerical result; dotted lines correspond to estimates by Eq. (\ref{eq:delta_phi}))}.}
    \label{fig3}
\end{figure*}

\section{Simulation results\label{sec3}}

Figure \ref{fig2} presents results of simulation of the electron hole dynamics performed for $A_0 = 2$. Panel (a) presents distributions of background magnetic field $B(s)$ and global electrostatic potential $\varphi(s)$ as a reference for other panels. The electric field and electrostatic potential of the electron hole at several moments of time are demonstrated in panels (b) and (c). The electron hole with $\Phi_0=0.8$ and $D=6$ is initially located in the neutral \blue{plane}, where $B(s)$ and $\varphi(s)$ are uniform. The electron hole has initial velocity $v_{H}=1.3$ and propagates into the region with $s > 0$. Panels (b) and (c) demonstrate that as the electron hole intrudes into and propagates in the region, where $B(s)$ and $\varphi(s)$ are non-uniform, the electron hole amplitude slightly grows, and bipolar electric field $E_{||}$ of the electron hole becomes asymmetric. In the electrostatic potential, the asymmetry comes out in appearance of a potential drop along the electron hole, which indicates the presence of a net electric field (double layer) localized around the electron hole. Panels (d)-(f) present the reduced distribution function, $F_{e}(s,v_{||},t)=\blue{2\pi}\int f_{e}(s,v_{||},\mu,t)\;B(s)d\mu$, at several moments of time and clearly demonstrate that propagation in the non-uniform plasma results in braking of the electron hole. At $t_{\rm ref}\approx 170$ the center of the electron hole is at $v_{||}\approx 0$ implying that the electron hole has been slowed down to $v_{H}\approx 0$. At that moment, the electron hole is reflected and then propagates back \blue{toward the neutral plane} (not shown). To evaluate effects of the global electrostatic potential $\varphi(s)$ on the electron hole dynamics, we have performed similar simulations for $A_0=1$ and 1.5.

\begin{figure*}[ht!]
    \centering
    \includegraphics[width=0.8\textwidth]{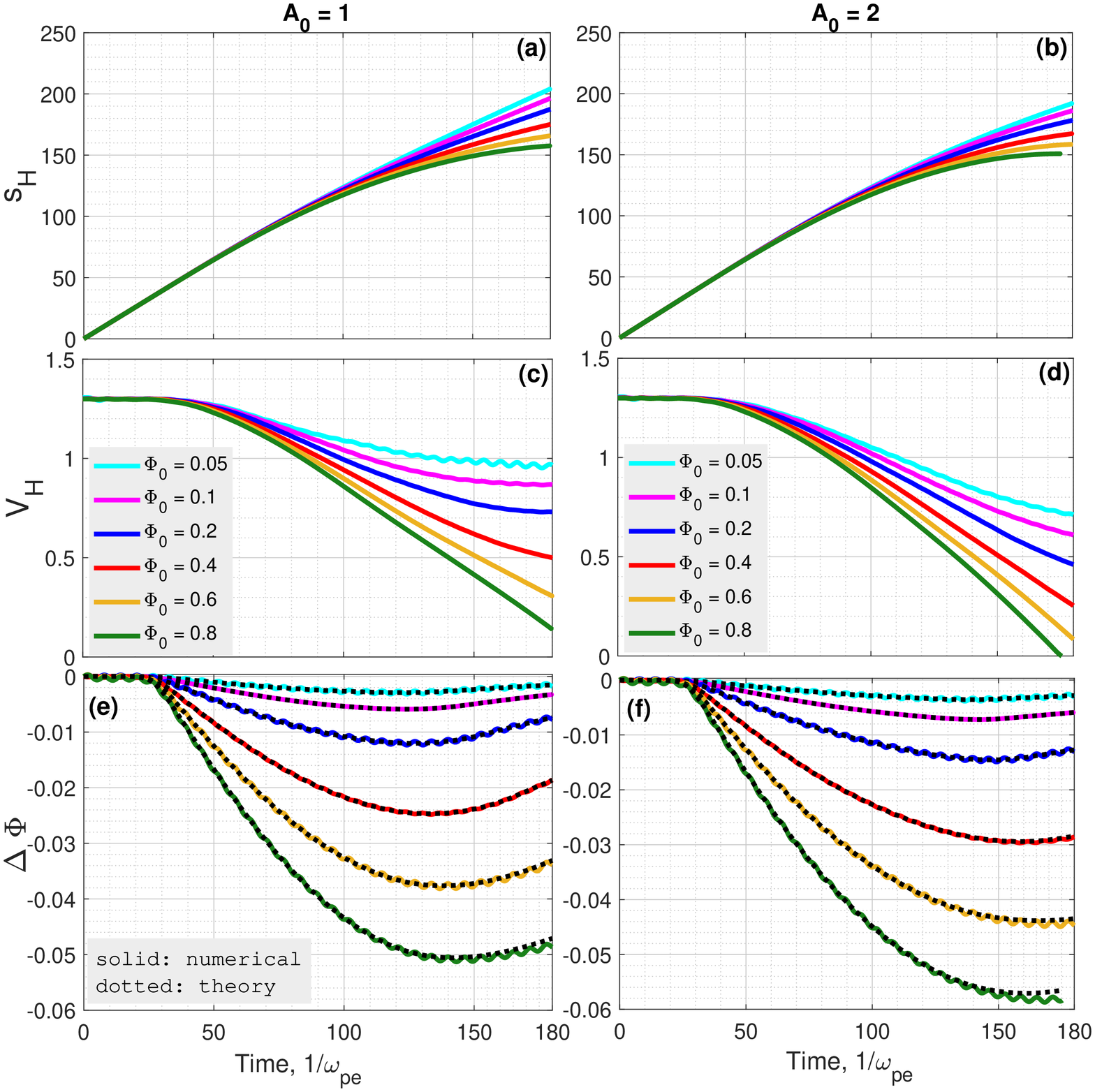}
    \caption{The results of simulations for electron holes with identical initial spatial width $D=6$ and velocity $v_{H}=1.3$, but different amplitudes $\Phi_0=0.05, ...,\, 0.8$. The simulations are performed for current sheets with electron temperature anisotropy $A_0=1$ and $A_0=2$. The panels present (a,b) electron holes trajectories $s_{H}(t)$ that is the position of the peak of electrostatic potential; (c,d) evolution of the electron hole velocities; (e,\blue{f}) evolution the potential drop along the electron hole; the solid curves indicate the potential drop in simulations, while dotted curves represent estimates given by Eq. (\ref{eq:delta_phi}).}
    \label{fig4}
\end{figure*}

Figure \ref{fig3} presents a summary of results of simulations for an electron hole with $\Phi_0=0.8$ and $D=6$ performed for various values of anisotropy parameter $A_0$. Panel (a) shows evolution of electron hole velocity $v_{H}=ds_{H}/dt$, where $s_H(t)$ is the position of the peak of the electron hole electrostatic potential $\Phi(s,t)$. In all simulations, the electron hole propagates with a constant velocity $v_{H}=1.3$ around the neutral \blue{plane}, where $B(s)$ and $\varphi(s)$ are uniform. As the electron hole propagates in the non-uniform plasma, its velocity decreases. \blue{In the simulation run for $A_0=1$, which correspond to $\varphi(s)= 0$, the electron hole significantly slows down at $t_{\rm ref}\approx 180$}, while in the simulation \blue{runs for $A_0>1$}, the electron holes are slowed down faster and reflected earlier. Thus, the presence of a non-zero global electrostatic potential $\varphi(s)$ results in faster braking and earlier reflection of electron holes, but the simulations also show that at typical values of electron \blue{temperature} anisotropy $A_0$ and global electrostatic potential $\varphi(s)$ the difference between the electron hole braking in simulations with $A_0=1$ \blue{and $1<A_{0}\lesssim 2$} is not crucial. In other words, the electron hole braking in the Earth's magnetotail current sheet is expected to be \blue{predominantly caused by} the non-uniform magnetic field. The braking of electron holes in a non-uniform plasma is similar to braking of electrons due to mirror force and global parallel electric field, but electron holes are actually slowed down \blue{at a different rate compared to individual electrons. In the  current sheet configuration under consideration, parallel velocity $\dot{s}$ of an electron is determined by the energy conservation law, ${\dot s}^2/2+\mu B(s)-\varphi(s)=C={\rm const}$, where $C=\dot{s}^2_0/2+\mu$ and $\dot{s}_0$ is electron velocity in the neutral plane. Panel (a) presents velocity evolution for individual electrons with $\dot{s}_{0}$ coinciding with initial electron hole velocity $\dot{s}_0=v_{H}(0)=1.3$ and magnetic field moment equal to averaged magnetic moment of electrons at the initial moment in the neutral plane, $\mu=1/A_0$. We can clearly see that electron hole with amplitude $\Phi_0=0.8$ slows down faster than individual electrons.}

Panels (b) and (c) show that the electron hole propagation in the non-uniform plasma results in a slight growth (within 10\%) of electron hole amplitude $\Phi_0(t)\equiv \Phi(s_H,t)$ and negligible variation of spatial width $D(t)$. The latter parameter is computed as the distance between minimum and maximum peaks of the bipolar electric field. Panel (d) demonstrates the potential drop along the electron hole computed as $\Delta \Phi = -\int_{-\infty}^{\infty}E_{||}ds$. The electron hole propagation in the non-uniform plasma results in a non-zero potential drop along the electron hole or, in other words, formation of a double layer localized around the electron hole. In fact, the potential drop can be estimated from the momentum conservation law for electrons (Section \ref{sec4}). The estimate of the potential drop is shown in panel (d) demonstrating a relatively good agreement with the simulation result. To determine parameters affecting evolution of velocity and potential drop of the electron hole we performed simulations for electron holes with various initial parameters. We considered electron holes with the same initial velocity $v_{H}=1.3$ and spatial width $D=6$, but assumed different initial amplitudes $\Phi_0$. Because the spatial width and amplitude do not strongly vary during the propagation, we demonstrate only evolution of electron hole velocities and potential drops.

Figure \ref{fig4} presents results of simulations performed for electron holes \blue{with different amplitudes $\Phi_0$} in a plasma with $\varphi(s)$ corresponding to $A_0=1$ and $A_0=2$. Panels (a) and (b) present electron hole trajectories $s_{H}(t)$, while panels (c) and (d) present evolution of electron hole velocities $v_{H}(t)$. These panels demonstrate that electron holes with larger amplitudes are slowed down faster. In other words, the \blue{effective mass-to-charge ratio} of electron holes propagating in a non-uniform \blue{decreases as amplitude $\Phi_0$ increases}. \blue{An analysis of the effective mass-to-charge ratio dependence on electron hole amplitude is given in the next section.} The comparison of trajectories and velocities of the electron holes in simulations with $A_0=1$ and $A_0=2$ confirms that although the global electrostatic potential contributes to the braking of electron holes in the current sheet, the leading contribution is provided by the non-uniform magnetic field. Panels (e) and (f) present evolution of the potential drop along the electron holes and show that the potential drop is larger for electron holes with larger amplitudes. That dependence of the potential drop on the electron hole amplitude is expected, based on the estimate of $\Delta \Phi$ presented in Section \ref{sec4}.

\begin{figure*}[ht!]
    \centering
    \includegraphics[width=0.8\textwidth]{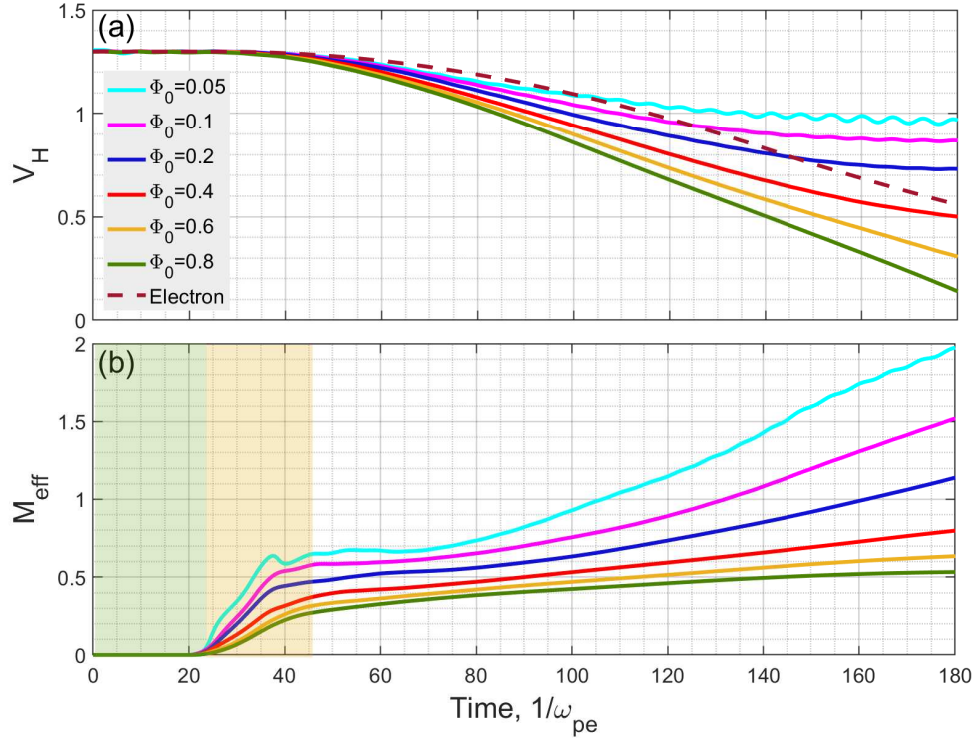}
    \caption{\blue{Analysis of the effective mass-to-charge ratio for electron holes with different amplitudes $\Phi_0$. We use results of the simulations carried out for $A_0=1$ (Figure \ref{fig4}). Panel (a) presents evolution of velocities of the electron holes and individual electrons with initial velocity in the neutral plane $\dot{s}_0=v_{H}(0)=1.3$ and magnetic moment $\mu=1$. Panel (b) presents the effective masses-to-charge ratios, $M_{eff}$, computed using Eq. (\ref{eq:Meff}), that is as the ratio between the local force and electron hole acceleration. The local force is zero, when the electron hole propagates in the region with uniform magnetic field and electrostatic potential, and, therefore, $M_{eff}$ could not be determined and set to zero for convenience (green-shaded region). Then follows the transition period (yellow-shaded), when the electron hole enters the non-uniform magnetic field, and $M_{eff}$ changes to its actual, slowly varying, value, which clearly depends on the electron hole amplitude. For larger amplitude electron holes, $M_{eff}$ slowly varies during the propagation, while for smaller amplitude electron holes, the initial period of almost constant $M_{eff}$ changes to a period with rather fast growth of $M_{eff}$. Thus, the concept of a constant effective mass-to-charge ratio has a limited applicability for small-amplitude electron holes.}}
    \label{fig5}
\end{figure*}

\section{Interpretation and discussion\label{sec4}}

We have demonstrated that electron holes produced by some instability around the neutral \blue{plane} and propagating to the current sheet boundaries are slowed. The simulations, including a global electrostatic field typical of the Earth's magnetotail current sheet, showed that though the electrostatic field does provide a noticeable contribution to the braking of electron holes, the leading contribution is provided by the non-uniform magnetic field. The electron holes are slowed in a similar fashion as individual electrons are slowed by a mirror force or parallel electric field. \blue{One could expect that electron holes behave like individual electrons and the electron hole dynamics could be described by the "energy conservation law",} $v_{H}^2(t)/2+\langle \mu\rangle B(s)-\varphi(s)={\rm C}$, where $\langle \mu\rangle=1/A_0$ is the averaged magnetic moment of trapped electrons \blue{at the initial moment} and $C=v_{H}^2(0)/2+\langle \mu\rangle$. \blue{However, the comparison of the dynamics of electron holes and individual electrons in Figure \ref{fig3} showed that electron holes are slowed at a different rate compared to individual electrons and, moreover, that rate should depend on the electron hole amplitude according to Figure \ref{fig4}}. Thus, to describe the dynamics of electron holes similarly to dynamics of individual electrons we need to assume some effective mass-to-charge ratio that depends on the electron hole amplitude. \blue{We compute the effective mass-to-charge ratio $M_{eff}$ as the ratio of a local force to a local electron hole acceleration rate} 
\begin{eqnarray}
\blue{M_{eff} = \dfrac{1}{dv_H/dt}\left(-\langle \mu\rangle\dfrac{\partial B}{\partial s} + \dfrac{\partial\varphi}{\partial s}\right)_{s=s_H(t)}},
\label{eq:Meff}
\end{eqnarray}
\blue{where $dv_{H}/dt$ is electron hole acceleration rate, the local force on the right-hand side is computed at $s=s_H(t)$ that is at the electron hole location, and $\langle \mu\rangle=1/A_0$ is the magnetic moment at the initial moment of time. We note that in chosen units the mass-to-charge ratio for individual electrons $M_{eff}=1$.}

\blue{Figure \ref{fig5} presents the evolution of the effective mass-to-charge ratio for electron holes with different amplitudes computed for simulations with $A_0=1$}. \blue{Panel (a) presents the evolution of electron hole velocities (already shown in Figure \ref{fig4}) along with evolution of velocity of individual electrons with magnetic moment $\mu=1$ and velocity of 1.3 in the neutral plane. The green shaded region in panel (b) indicates the time interval, where the magnetic field and electrostatic potential were uniform and, hence, there was no force acting on electrons. In that region, the effective mass-to-charge ratio cannot be calculated using Eq. (\ref{eq:Meff}), because both the local force and acceleration rate are zero. The effective mass-to-charge ratio can be determined only as the electron hole enters and propagates in the region with non-uniform magnetic field and electrostatic potential. After a short transition period indicated in panel (b), $M_{eff}$ approaches some value that varies slowly in time and can be considered as the effective mass-to-charge ratio of the electron hole. One can see that the mass-to-charge ratio depends on the electron hole amplitude: electron holes with larger amplitudes have lower effective mass-to-charge ratios, and the dynamics of such electron holes deviates stronger from the dynamics of individual electrons, i.e. $M_{eff}$ is rather different from 1. Electron holes with smaller amplitudes have larger effective mass-to-charge ratio with values closer to 1. Another interesting feature that can be noticed in panel (b) is that for larger amplitude electron holes, the effective mass-to-charge ratio does not significantly vary during the electron hole propagation, so the notion of the effective mass-to-charge ratio is meaningful. On the other hand, for small-amplitude electron holes, the initial period of almost constant effective mass-to-charge ratio changes to a rather rapid growth. This is likely due to processes of electron trapping and detrapping, which results in variation of the averaged magnetic moment $\langle \mu\rangle$ in Eq. (\ref{eq:Meff}). The processes of trapping and detrapping occur due to the mirror force acting on trapped electrons\cite{Vasko16:pop} and, therefore, are expected to stronger affect the dynamics of electron holes with smaller amplitudes.} 

\blue{Although the quasi-particle concept does not perfectly describe the dynamics of electron holes ({\it ad-hoc} effective mass-to charge ratio varying in time should be included to reproduce the simulations results), that naive concept allows qualitative understanding} of the presented simulation results. In particular, according to this concept the dominant contribution into the electron hole braking in the current sheet should be indeed due to the non-uniform magnetic field, because typical electrostatic field  $\partial \varphi/\partial s$ is a few times smaller than mirror force $\langle \mu\rangle \partial B(s)/\partial s$ (Figure \ref{fig1} shows that $\partial \varphi/\partial s\sim 10^{-3}$, while $\partial B(s)/\partial s\sim 10^{-2}$).

We have shown that the spatial width of electron holes does not substantially vary in the course of propagation. This is consistent with 1D Vlasov simulations of electron holes in a plasma with non-uniform plasma density $n_0(s)$ (see Ref. \cite{Vasko17:pop}) that showed that the spatial width varies as $D(t) = D(0)n_0^{-\nu}(s_H)$, where positive index $\nu$ depends on the the electron hole amplitude. The simulations presented in this study demonstrate that evolution of the spatial width of electron holes is solely dependent on the plasma density, and not on spatial distributions of $B(s)$ or $\varphi(s)$. The simulations have also shown that the electron hole amplitude does not substantially vary in the course of propagation, which has to be addressed analytically in future studies.

The simulations demonstrated that propagation in a non-uniform plasma results in formation of a potential drop along the electron hole or, in other words, a double layer localized around the electron hole. The necessity of formation of the potential drop can be deduced from the electron momentum equation. Multiplying the Vlasov equation (\ref{eq:ampere}) by $v_{||}$ and integrating over $B(s)dv_{||}d\mu$ we obtain the momentum equation for electrons
\begin{eqnarray*}
-\frac{\partial j_{||}}{\partial t}+\frac{\partial P_{||}}{\partial s}+\frac{P_{\perp}-P_{||}}{B}\frac{\partial B}{\partial s}-\left(\frac{\partial \varphi}{\partial s}-E_{||}\right) n_{e}=0
\end{eqnarray*}
where $P_{||}$ and $P_{\perp}$ are electron pressures
\begin{eqnarray*}
P_{||} = \int v_{||}^2\;f_{e}\;B(s)\;dv_{||}d\mu,\;\;\;\;
P_\perp = \frac{1}{2}\int v_{\perp}^2\;f_{e}\;B(s)\;dv_{||}d\mu
\end{eqnarray*}
while $n_{e}$ and $j_{||}$ are electron density and current density, which are given by Amp\'{e}re and Poisson equations
\begin{eqnarray*}
j_{||} = -\frac{\partial E_{||}}{\partial t}, \;\;\;\;\;n_{e}=1+B(s)\frac{\partial }{\partial s}\left[\frac{1}{B(s)}\left(\frac{\partial \varphi}{\partial s}-E_{
||}\right)\right]
\end{eqnarray*}
Integrating the momentum equation over the simulations box we obtain
\begin{eqnarray}
\label{eq:delta_phi}
    \Delta\Phi = -\int_{-s_{\rm max}}^{s_{\rm max}}\left[P_{||}-P_{\perp}-\left(\frac{\partial \varphi}{\partial s}-E_{||}\right)^2\right]\dfrac{\partial\ln{B}}{\partial s}ds,
\end{eqnarray}
where we have neglected $d^2\left(\Delta \Phi\right)/dt^2$ on the left-hand side, because in a weakly non-uniform plasma $\Delta \Phi$ varies slowly in time, so that $d^2\left(\Delta \Phi\right)/dt^2\ll \Delta \Phi$. As compared to Ref. \cite{Kuzichev17:grl}, the additional term is $\partial \varphi/\partial s$ on the right-hand side. Eq. (\ref{eq:delta_phi}) has been used to estimate $\Delta \Phi$ shown in Figures \ref{fig3} and \ref{fig4}, where $P_{||}$, $P_{\perp}$ and $E_{||}$ were adopted from the simulations. Eq. (\ref{eq:delta_phi}) can be simplified to derive rather simple expression for the potential drop to be used for the order of magnitude estimates. First, the integrand is localized around the electron hole, that is why the potential drop is accumulated around the electron hole (Figure \ref{fig2}). Second, $P_{||}$ and $P_{\perp}$ can be replaced by their local variations $\delta P_{||}$ and $\delta P_{\perp}$ arising and localized around the electron hole. Third, estimates by the order of magnitude show that $\partial \varphi/\partial s\ll E_{||}$ (Figure \ref{fig1}) and $E_{||}^2\ll \delta P_{||}$, because $\delta P_{||}\sim \Phi_0$ and $E_{||}^2\sim \Phi_0^2/D^2$, where $D$ is of the order of a few, while $\Phi$ is typically below a few tenths (see, e.g., Ref. \cite{Lotekar20}). Therefore, the potential drop can be estimated by the order of magnitude as follows
\begin{eqnarray}
\Delta \Phi\sim -(\delta P_{||}-\delta P_{\perp})\;D\;\left[\frac{1}{B}\frac{\partial B}{\partial s}\right]_{s=s_H}
\label{eq:simple}
\end{eqnarray}
The presence of electron hole locally increases the parallel pressure resulting in some positive $\delta P_{||}-\delta P_{\perp}$ that is larger for larger electron hole amplitudes, because $\delta P_{||}\sim \Phi_0$. Therefore, the potential drop is larger for electron holes with larger amplitudes in accordance with the simulation results. In addition, Eq. (\ref{eq:simple}) shows that in the realistic Earth's magnetotail current sheet the potential drop will be extremely difficult to measure, because $D\ll L\equiv B(\partial B/\partial s)^{-1}$, where $L$ is typical scale of magnetic field variation.

Finally, we discuss caveats and applications of the presented simulations. \blue{First,} the ion dynamics was not included into the simulations that is reasonable while electron hole velocities are larger than typical ion thermal velocity. 
Once the electron hole velocity becomes comparable to typical ion thermal velocity, the interaction with ions via Landau resonance can strongly affect the electron hole evolution\cite{Zhou16,Lotekar20}. 
Second, the presented simulations have only one spatial dimension and do not take into account instabilities destroying the coherency of electron holes and restricting electron hole lifetimes. 
\blue{The modern multi-dimensional simulations showed that electron holes can survive for up to a few hundred of electron plasma periods \cite{Goldman99,Oppenheim01:grl}, and spacecraft measurements support these estimates \cite{Norgren15,Lotekar20}. 
Therefore, in realistic situations, magnetic field gradients will affect the dynamics of electron holes if the spatial scale of the magnetic field variation is smaller than about $10^3\;v_{H}/\omega_{p}\sim 10^3\;\lambda_{D}$, 
where we have assumed $v_{H}$ to be of comparable to the electron thermal velocity. In the Earth's magnetotail $\lambda_{D}\sim 1$ km and, hence, magnetic fields with spatial scales of less than about one thousand kilometers can affect the dynamics of electron holes. 
Thirdly, in the present simulations we do not model an instability producing the electron holes. Instead, we consider a smooth laminar situation to consider a particular physical effect rather than to reproduce all stages of development of a particular instability in a 
non-uniform plasma. Interestingly, the effect of acceleration of electron holes in the course of propagation toward the current sheet neutral plane was reported in fully self-consistent Particle-In-Cell (PIC) simulations of the magnetic reconnection in the Earth's 
magnetotail\cite{Goldman14:prl}. Thus, although the electron hole dynamics in PIC simulations is superimposed on turbulent and noisy fluctuations, still the effect of electron hole propagation in a non-uniform plasma was noticeable. In addition, the simulations in Ref. 
\cite{Goldman14:prl} demonstrate that magnetic field gradients typical of reconnecting current sheets in the Earth's magnetotail can indeed affect the dynamics of electron holes produced during the magnetic reconnection.}\newline
\indent Among useful applications we point out \blue{additional scenarios} of the origin of slow electron holes recently reported in current sheets in the Earth' magnetotail\cite{Norgren15,Lotekar20} and at the magnetopause\cite{Graham15,Graham16}. It was suggested that these electron holes are \blue{likely produced locally by the Buneman instability \cite{Drake03,Che09:prl,Che10}. Although that is very likely the case, there have not been presented direct experimental evidence supporting this point of view. To a large extent, the lack of direct evidence is due to rather fast relaxation of the Buneman instability, so that plasma instruments aboard modern spacecraft do not resolve the instability development, but instead measure only a saturated state\cite{Norgren15,Lotekar20}. Nevertheless, one cannot rule out that some slow electron holes can} in principle appear due to slowing down of originally fast electron holes produced by electron beam instabilities around the neutral \blue{plane} and slowed down in the course of propagation in the current sheet. That scenario is viable provided the electron hole survive over sufficiently long time (as already discussed above), which is not unrealistic, because there were reports of electron holes measured sequentially aboard several spacecraft separated in space by a few tens of kilometers \cite{Pickett04:npg,Norgren15,Lotekar20}. \blue{We also note that direct evidence for electron holes produced by an electron beam instability in the Earth's magnetotail current sheets has been recently provided in Refs. \cite{Holmes18,Tong18}}.

\section{Conclusion\label{sec5}}

We have presented 1.5D Vlasov simulations of the electron hole dynamics in a non-uniform plasma typical of current sheets in the Earth's magnetosphere and, particularly, in the Earth's magnetotail. The results of the study can be summarized as follows
\begin{enumerate}
    \item the global electrostatic field, which is typically present in current sheets, contributes to braking of electron holes propagating toward the current sheet boundaries; however, the dominant contribution into the electron hole braking is provided by the non-uniform magnetic field.
    \item the braking of electron holes occurs slower for electron holes with smaller amplitudes; \blue{thus, the effective mass-to-charge ratio is smaller for electron holes with larger amplitudes.}
    \item the propagation in a current sheet results in appearance of the potential drop along the electron hole that is larger for electron holes with larger amplitudes.
    \item the spatial width and amplitude of electron holes do not substantially vary in the course of propagation in current sheets.
\end{enumerate}

\section*{Acknowledgments}
The work of P.S. and A.A. was supported by the Russian Science Foundation grant No. 19-12-00313. The work of I.V. was supported by NASA grant 80NSSC19K1063 and NASA Heliophysics Supporting Research grant 80NSSC20K1325. The work of I.K. was supported by NSF Grant AGS-1502923 and the NASA Van Allen Probes RBSPICE instrument project provided by JHU/APL Subcontract No. 131803 to NJIT under NASA Prime Contract No. NNN06AA01C. I.K. would like to thank Dr. Gareth Perry for providing an access to his computational resources at NJIT Lochness HPC cluster.

\section*{Data availability}
The data that support the findings of this study are available from the corresponding author upon reasonable request.






\input{aipsamp.bbl}

\end{document}

%% file: aipsamp.bbl
%